\documentclass[12pt]{article}
\usepackage{latexsym}
\usepackage[T1]{fontenc}

\usepackage[pdftex,dvips]{graphicx}
\usepackage{array} 
\usepackage{amssymb,amsfonts,amsmath}
\usepackage{color}

\usepackage{amssymb,amsfonts,amsmath}
\usepackage{color}

\newenvironment{sciabstract}{%
\begin{quote}
\bf
\end{quote}}

\begin{document}
\baselineskip24pt

{\Large Classification: Physical Sciences - Applied Physical Sciences - Biological Sciences - Biophysics and Computational Biology}\\
\\
{ \center{\bf \LARGE A new look at blood shear-thinning}\\
Luca Lanotte$^{1}$, Johannes Mauer$^{2}$, Simon Mendez$^{3}$, Dmitry A. Fedosov$^{2}$, Jean-Marc Fromental$^{4}$, Viviana Claveria$^{1}$, Franck Nicoud$^{3}$, Gerhard Gompper$^{2}$, and Manouk Abkarian$^{1}$\\

\noindent Authors affiliation: \\
$^{1}${Centre de Biochimie Structurale, CNRS UMR 5048 - INSERM UMR 1054, University of Montpellier,34090 France}\\
$^{2}${Institute of Complex Systems and Institute for Advanced Simulation, Forschungszentrum J\"ulich, 52425 Germany}\\
$^{3}${IMAG UMR 5149 CC 051, University of Montpellier, 34095 Montpellier}\\
$^{4}${Laboratoire Charles Coulomb, CNRS UMR 5587, University of Montpellier,34095 France}

\noindent Corresponding author: \\
Manouk ABKARIAN\\
Email: Manouk.Abkarian@umontpellier.fr\\
}


\title{A new look at blood shear-thinning} 





\maketitle

\begin{sciabstract}
  \bf  \color{red}{Blood viscosity decreases with shear stress, a property essential for an efficient perfusion of the vascular tree. Shear-thinning is intimately related to the dynamics and mutual interactions of red blood cells (RBCs), the major constituents of blood. Our work explores RBCs dynamics under physiologically relevant conditions of flow strength, outer fluid viscosity and volume fraction. Our results contradict the current paradigm stating that RBCs should align and elongate in the flow direction thanks to their membrane circulation around their center of mass, reducing flow-lines disturbances. On the contrary, we observe both experimentally and with simulations, rich morphological transitions that relate to global blood rheology. For increasing shear stresses, RBCs successively tumble, roll, deform into rolling stomatocytes and finally adopt highly deformed and polylobed shapes even for semi-dilute volume fractions analogous to microcirculatory values. Our study suggests that any pathological change in plasma composition, RBCs cytosol viscosity or membrane mechanical properties will impact the onset of shape transitions and should play a central role in pathological blood rheology and flow behavior.}
 \end{sciabstract}

%


\maketitle 









\section*{Significance Statement}
{\bf Blood viscosity decreases with shear stress, a property essential for an efficient perfusion of the human body. This shear-thinning property is intimately related to the dynamics of its major constituents, the red blood cells (RBCs). Our work explores this dynamics under physiologically relevant conditions of both flow and viscosity and contradicts the current paradigm describing RBCs behavior as the one of viscous droplets. Using both experiments and simulations, our study shows that the lack of membrane fluidity is the key feature which controls dynamic morphological transitions and global rheology. Our study questions the physiological relevance of previous droplet-like descriptions of RBCs dynamics.}

\section{Introduction}
Red blood cells (RBCs) are the main cellular component of whole blood (WB). About thousand times more concentrated
than white blood cells and platelets, \textcolor{red}{the volume fraction of RBCs, also called the hematocrit (Ht), is as high as $45\%$.
Local dynamics and interactions of RBCs govern blood rheology and control proper perfusion of the entire vascular tree. In particular, WB exhibits a strong shear-thinning behavior
in shear flow \cite{Wells:1961vm,Dintenfass:1968ta} which is determined by the
aggregability and deformability of RBCs \cite{Chien:1970wz,Skalak_MBF_1981}, since the suspending plasma behaves essentially as a Newtonian fluid with a constant shear viscosity
of about $1.2$ cP at $37^{\circ}$C \cite{Wells:1961vm,Brust:2013ia}}. Macromolecules dispersed in the plasma, such as fibrinogen, induce an attractive
force between RBCs \textcolor{red}{leading to} their aggregation \cite{Merrill_BRE_1966,Wagner:2013bw}. Because RBCs have a biconcave disk-like shape at rest, they form
long floppy rouleaux structures which can reversibly and continuously break down to single flowing discocytes for increasing
shear rates $\dot{\gamma}$ up to tens of s$^{-1}$ \cite{Qin_ESF_1998,Fedosov_PBV_2011}. This change in microstructure is indeed 
accompanied by a strong viscosity drop from $\sim100$ cP down to about $\sim10$ cP \cite{SchmidSchonbein:1971vd}. However, for shear 
rates between $10$ and $1500$ s$^{-1}$, which are very common in the microcirculation \cite{Lipowsky:2005,Popel:2005ih}, 
only the deformability and dynamics of RBCs can account for a further 4-fold decrease in WB viscosity down to values as low 
as $2-3$ cP \cite{Chien:1970wz,Skalak_MBF_1981}. This value strongly depends on the hematocrit and the viscosity of the hemoglobin cytoplasm of 
RBCs, which is around $5$ cP at $37^{\circ}$C \cite{Wells:1969ww}.

The link between cellular deformability and shear-thinning remains unsettled, even though it is crucial in understanding blood flow
both in health and disease, since altered deformability of RBCs \cite{Diez_Silva_SBM_2010} has indeed been correlated with impaired 
perfusion, increase in blood viscosity, and vaso-occlusion in microcirculatory disorders such as diabetus mellitus, or 
hemoglobinopathies like sickle cell anemia. Measurements of WB rheology in microtubes \cite{Skalak_MBF_1981,Goldsmith:1972up} 
and rheoscopes \cite{SchmidSchoenbein:1969tz,Fischer:1978wb},
realized during \textcolor{red}{the 1970s and the 1980s}, have led to the foundation of the current paradigm for shear-thinning, comforting an emulsion
analogy \textcolor{red}{inherited from the 1960s} \cite{Wells:1961vm,Dintenfass:1968ta}. While at small shear stresses (less than about $5\times10^{-2}$ Pa),
single RBCs flip like a coin, they attain a steady orientation 
with respect to the direction of the flow for increasing shear stresses \cite{SchmidSchoenbein:1969tz,Fischer:1978wb}. 
\textcolor{red}{This concept is correct only when RBCs are dispersed into a viscous aqueous 
  solution (often of dextran polysaccharides), which is several times more viscous than the inner cytoplasm, as has been done in most of available experiments} \cite{SchmidSchoenbein:1969tz,Fischer:1978wb}.
Such a viscous environment is supposed 
to mimic the effective high-shear-rate viscosity of the WB suspension measured by rheometry \cite{Fischer:1977vw}. Single RBCs 
in such viscous solutions behave similar to liquid droplets. Their composite membrane elements rotate around their center of mass as the tread of a tank, transferring to the cytoplasm the vorticity necessary for orientation stabilization \cite{Fischer:1978wb}. Membrane 
tank-treading with a steady alignment would effectively assist the flow by minimizing the disturbances to flow lines. 
Subsequent gradual elongation of the discocytes into prolate ellipsoids for shear stresses increasing up to few Pa would 
then explain \textcolor{red}{the deformation- and shape-controlled shear-thinning behavior of WB} \cite{Fedosov_PBV_2011}. Therefore, an essential element of this paradigm 
is that single RBCs \textcolor{red}{should} tank-tread in response to high shear stresses.

Recent experiments \textcolor{red}{on dilute RBC suspensions}, realized with outer viscosities similar to that of the hemoglobin-rich cytoplasm, have demonstrated however that the solid-like tumbling motion of isolated RBCs is not replaced by the tank-treading dynamics for increasing shear rates, but rather by a typical rolling motion, where the axis of symmetry of the discocyte lies in the direction of the vorticity \cite{Dupire:2012cx}. Even for shear stresses as high as $0.1-0.5$ Pa, no fluidization of the membrane was observed, demonstrating the important role the inner-to-outer viscosity ratio $\lambda$ plays in local dynamics. In fact, the viscosity of blood plasma under physiological conditions is about five times smaller than that of the RBC cytosol. Hence, new studies are required to explore the local dynamics of RBCs at such low viscosity conditions, especially when $\dot{\gamma}$ is greater than $10$ s$^{-1}$ or in the range of stresses between a tenth to a few Pa, which are characteristic for microcirculatory flow \cite{Lipowsky:2005,Popel:2005ih}.

\textcolor{red}{In this work, we explore the shape and dynamics of RBCs in the range of shear rates and hematocrits relevant for the microcirculation and correlate local RBCs morphologies with global shear-thinning behavior measured by rheometry. Using microfluidics, we demonstrate that RBCs show unexpected dynamics characterized by rotating polylobed shapes, with no clear tank-treading of their membrane. Experimental results obtained by both microfluidic observations and rheological measurements are corroborated by two different 3D simulation techniques: a continuum approach YALES2BIO \cite{Mendez:2014a} based on the finite-volume method and 
a mesoscopic approach for blood-flow modeling \cite{Fedosov_RBC_2010} based on the smoothed dissipative particle dynamics (SDPD) method \cite{Mueller_SDPD_2015} (\textit{see Materials and Methods}). We discuss the implication of this work for microcirculatory flow and re-interpret the classical shear-thinning conjecture.}


\section{Results}
\subsection{Morphology of RBCs in simple shear flow} 
\textcolor{red}{We explore first the morphologies of single RBCs at $37^\circ$C submitted to a simple shear flow in a range of $\dot{\gamma}$ between $10$ s$^{-1}$ and $2000$ s$^{-1}$, using a cone-and-plate configuration of a rheometer. Cells are hardened with an aldehyde treatment while flowing (see protocol in Materials and Methods)}. Samples of these hardened RBCs (HRBCs) are then withdrawn from the plate of the rheometer and visualized by bright field (BF) and confocal microscopy after fluorescent dying of their membranes. Typical images of the fixed shapes are shown in fig. \ref{fig1}A \textcolor{red}{for a given sample}. We classify the different morphologies of HRBCs as a function of $\dot{\gamma}$ in fig. \ref{fig1}B using an in-house image-analysis program.

RBCs subjected to weak shear rates ($<10$ s$^{-1}$) commonly keep their biconcavity similar to typical discocytes at rest (fig. \ref{fig1}A). For $\dot{\gamma}$
increasing from $10$ to $40$ s$^{-1}$ however, the percentage of discocytes decreases more than twice (fig. \ref{fig1}B). This is associated
with the systematic development of two populations of cup-shaped stomatocytes: one with a circular rim and another with an elliptical one \textcolor{red}{ with an aspect ratio smaller than 0.95}. Examples of these
deformed stomatocytes are displayed in fig. \ref{fig1}A at $18$ and $40$ s$^{-1}$, respectively. At $\dot{\gamma}=45$ s$^{-1}$, the fraction of discocytes
has dramatically dropped representing only about 30\% of the total population, while the total amount of stomatocytes has jumped nearly to $65\%$ of the sample.
These values remain nearly constant for both cell populations up to about 400 s$^{-1}$ (light blue color region of fig. \ref{fig1}A). However, in the region between
$45$ s$^{-1}$  and $400$ s$^{-1}$, the fraction of circular-rim-shaped stomatocytes decreases, and gets gradually replaced by the population of elliptical-rim-shaped
stomatocytes with an increased ellipticity. 

At $\dot{\gamma} \approx 400$ s$^{-1}$, a new substantial transition takes place. As the shear rate goes beyond $400$ s$^{-1}$, we observe a sharp increase
in the population of deformed stomatocytes which appear to be strongly folded (image for $475$ s$^{-1}$ in fig. \ref{fig1}A). Most importantly, we also
see the rise of a new population of RBCs showing large lobes on their surface. In most cases, we find cells with three or six lobes forming tetrahedra,
which will be henceforth referred to as trilobes or hexalobes, respectively. In the region $400$ s$^{-1}<\dot{\gamma}<2000$ s$^{-1}$, marked by a light red
color in fig. \ref{fig1}B, we observe the nearly disappearance of discocytes (less than 10\%) and the formation of trilobes and to a lesser degree hexalobes.
At $\dot{\gamma}= 2000$ s$^{-1}$, the polylobed RBCs represent almost 70\% of the sheared population. Moreover, rapid rise of hexalobe population is found for
$\dot{\gamma}\sim2000 $ s$^{-1}$, suggesting a direct correlation between the number of lobes and shear strength. 

\subsection{RBCs dynamics in flow} 

\textcolor{red}{To understand further how the acquired morphologies couple to the flow and to rule out hardening
artifacts, we perform complementary microfluidic experiments with high-speed video microscopy at $25^\circ$C using dilute suspensions of RBCs ($Ht\approx1\%$)  in PBS/BSA only (no hardening). The suspension is fed at increasing flow rates into a circular cross-section microcapillary for the same range of shear rates explored previously by rheometry. Local shear rate is evaluated by measuring both the local cell velocity and its distance from the capillary walls. Different dynamics are detected for increasing
values of this estimated $\dot{\gamma}$}. Their time sequences are  shown in fig. \ref{fig2}A obtained in a field of view of about $300$ $\mu$m long. At low shear
rates ($\dot{\gamma}<40$ s$^{-1}$), the tumbling-to-rolling transition is observed, where the axis of symmetry of discocytes gradually aligns with
the vorticity direction. As in the rheological setup, the number of rolling RBCs becomes significant for increasing shear rates. In addition,
a substantial increase in the population of cup-shaped stomatocytes is clearly detected, which show a rolling motion in the flow as illustrated in fig. \ref{fig2}A at $\dot{\gamma}=150$ s$^{-1}$.
Moreover, observations reveal both tumbling and vacillating-breathing behaviors (see Movie S1) of folded stomatocytes (fig. \ref{fig2}A, blue box) with the latter being largely
detected at high shear strengths $\dot{\gamma}>350$ s$^{-1}$ (see examples in fig. \ref{fig2} for $250$ s$^{-1}$and $500$ s$^{-1}$). Furthermore, these \textit{in vitro} investigations confirm the formation and stability of polylobed shapes at higher shear rates. Two trilobes are depicted in side and top views at $750$ and $800$ s$^{-1}$,
respectively, in fig. \ref{fig2}A (light red color box) and in Movie S2. At such high shear rates, RBCs elongate in the vorticity direction, and display three lobes which
rotate around the center of mass. An even more solid-like rotation is observed for hexalobes under comparable flow conditions (last time sequence
in fig. \ref{fig2}A and in Movie S3). These experiments show that no sharp morphological transition occurs for increasing shear rates, but rather a marked variation
in shape distribution of RBC populations is present. Finally, thanks to our pressure injection system, we can produce very fast ``stop flow'' experiments.
These experiments nicely demonstrate the dynamical nature of polylobed morphologies, since either trilobes or hexalobes return rapidly to their resting
discocyte shape, when the flow is stopped, as depicted in fig. \ref{fig2}B and in Movie S4. Noteworthy, the population of stomatocytes keeps its shape longer and relaxes
to the discocyte population on a longer time scale of tens of minutes.

To complement the experiments, we also perform numerical simulations of isolated RBCs in simple shear flow for varying shear rates, using both SDPD and 
YALES2BIO softwares \textcolor{red}{(\textit{see Materials and Methods})}. The simulated dynamics is presented in fig. \ref{fig2}A using only YALES2BIO results, since agreement between the two simulations  is very good (some supplementary movies in SDPD are also provided). Simulations confirm the sequence of shapes and dynamics observed experimentally as well as their dependence on $\dot{\gamma}$ 
as shown in fig. \ref{fig2}A. In comparison to experiments, simulations provide a detailed information about dynamic behavior of RBCs in flow. 
For example, the same cell may have different stable shapes under flow (observed during several hundreds of time units $1/\dot{\gamma}$) depending on 
its initial orientation with respect to the flow. In particular, stomatocytes can display both rolling-like or tumbling-like motion; rolling deformed stomatocytes (see Movie S5) and tumbling trilobes (see Movie S6) might be found for the same shear rate. However, the sequence of shapes found experimentally appears to be qualitatively 
robust to moderate changes in the RBC mechanical properties, which has been tested in simulations. One difference between experimental and simulation 
results is hexalobe deformation of RBCs (fig. \ref{fig2}A at the bottom), which has been observed in simulations only as a transient state. 
Nevertheless, in none of the simulations performed for the range $1$ s$^{-1}$$<\dot{\gamma}<2000$ s$^{-1}$ a tank-treading ellipsoidal RBC has been obtained.

\subsection{Morphology and dynamics of RBCs at high hematocrits}

To evaluate the prevalence of polylobed shapes at physiologically relevant shear rates and hematocrits, same hardening experiments done in flow and described earlier 
were realized for several RBC suspensions with different $Ht$ values of 5, 15, 22, 35 and 45\%. $\dot{\gamma}=900$ s$^{-1}$ (at 37$^\circ$C) 
has been selected for these experiments, since the analysis for dilute suspensions in fig. \ref{fig1}B has shown a maximum in the population 
of trilobes at this shear rate. Figure \ref{fig3}A indicates a significant decrease in the \textcolor{red}{probability to obtain polylobed HRBCs} for increasing 
$Ht$. The percentage of trilobes and hexalobes, which amounts for about $55\%$ of the entire population at $Ht = 5\%$, is more than halved 
when the hematocrit reaches $45\%$. In addition, at high enough $Ht$ many HRBCs display a deformed shape with multiple irregular lobes, 
making it difficult to precisely classify their morphology; these cells will be called ``multilobes'' with an illustration in the lower inset of 
fig. \ref{fig3}A. The fraction of these  ``multilobed'' RBCs  seems to be rather stable amounting for $25\%-35\%$ of the total population. 
However, the onset of a new cell morphology is detected for increasing $Ht$. Flattened discocytes characterized by one or several grooves 
or creases on their membrane have been found and are illustrated in the inset of fig. \ref{fig3}A by optical and confocal microscopy. 
The fraction of creased discocytes increases significantly at high $Ht$ values amounting for almost $50\%$ of the sample at $Ht =45\%$. 
In contrast to the vast majority of polylobed RBCs in dilute suspensions, the formation of flat cells with creases is a clear effect 
of mutual hydrodynamic interactions between cells in such dense suspensions. Indeed, the total percentage of lobed cells with increasing $Ht$ 
has the exact inverse evolution of that for creased discocytes as shown in fig.\ref{fig3}A. Interestingly, physiological hematocrits in 
the microcirculation lie in the range $15-30$\% (light yellow region in fig.\ref{fig3}A), in which a clear coexistence of different 
cell morphologies with a domination of lobed shapes is present.

To corroborate the existence of such folded and irregular structures in flow at high $Ht$, we have performed SDPD simulations with dense RBCs 
suspensions. The analysis of creased discocytes population for different $Ht$ is presented in fig. \ref{fig3}A along with the corresponding 
experimental data showing a good qualitative agreement. Figure \ref{fig3}B presents RBC-extension distributions for different $Ht$ values,
which are averaged over multiple RBCs and many time instances (A typical simulation is shown in Movie S7). The RBC extension corresponds to the maximum instantaneous cell length in the flow direction, which is illustrated 
in one of the insets in fig. \ref{fig3}B. The extension distributions clearly show that RBCs become more stretched in the flow as $Ht$ is increased. The two regions indicated by light yellow and light blue colors in fig. \ref{fig3}B schematically illustrate 
the ranges of cell extensions with polylobed shapes and creased discocytes. Thus, at low enough $Ht$ polylobed shapes dominate, while at $Ht=45\%$ the fraction of creased discocytes becomes larger than one half. The coexistence of various cell morphologies and the evolution of different RBCs populations with increasing $Ht$ are fully consistent with the observations made experimentally.       

\subsection{The role of cell deformability in shear-thinning rheograms}
Now, we re-examine the interpretation of classical shear-thinning rheograms for human blood, which were introduced in the 1970s \cite{Chien:1970wz} and provided the seminal link between \textcolor{red}{shear-induced RBC deformation, tank-treading}, and shear-thinning. We perform experiments at $Ht=45\%$ suspending RBCs in either PBS/BSA or their native plasma. The samples are sheared in the rheometer at 37$^{\circ}$C (using both Couette and cone-and-plate geometry) for the range of $\dot{\gamma}$ between $5$ s$^{-1}$ and $2000$ s$^{-1}$. The relative viscosity ($\eta_{rel}$),defined as the ratio between the measured viscosity of the suspension ($\eta_s$) and the viscosity of the suspending medium ($\eta_m$), is obtained for different $\dot{\gamma}$. Consistently with previous works \cite{Chien:1970wz}, at low shear rates, washed blood has a significantly lower viscosity than WB since plasma induces aggregation. From $\dot{\gamma}\approx10$ s$^{-1}$, viscosity 
curves for both washed blood and WB merge into a common shear-thinning line as shown in fig. \ref{fig4}A. To understand the role of 
shape change in this regime, we compare washed blood rheology with that of two HRBC suspensions at 45\% $Ht$: one with stiff discocytes hardened at rest and the other one with polylobed cells obtained by hardening cells \textcolor[rgb]{1,0,0}{previously in the rheometer at $\dot{\gamma}=1500$ s$^{-1}$ and resuspending them at the desired 45\% $Ht$.
Both suspensions of HRBCs show a nearly Newtonian behavior except for shear rates above 400 s$^{-1}$ where the onset of a slight shear-thickening is observed especially for hardened discocytes. This behavior is similar to shear-thickening reported for rigid colloidal suspensions \cite{Barnes:1989,Brown:2011}. For both HRBC dispersions viscosity values are larger than those for washed blood}; however, the sample with hardened polylobes yields a $70\%$ lower relative viscosity than that for stiffened discocytes, as seen in Fig \ref{fig4}A. These results indicate that from $\dot{\gamma}=10$ s$^{-1}$ up to a few hundred s$^{-1}$, shear-thinning is largely determined by the change in RBC dynamics from tumbling to rolling and by the deformation of RBCs into elongated stomatocytes. Beyond $\dot{\gamma}\sim400$ s$^{-1}$ however, 
a further viscosity drop is related to the formation of polylobed and flattened shapes discussed above. 

SDPD simulations performed for $\lambda=5$ and $Ht=45\%$ quantitatively capture shear-thinning as shown 
in fig. \ref{fig4}A. A closer look at RBCs shapes for different shear rates essentially confirms experimental 
observations. For moderate shear rates up to a few hundred s$^{-1}$, a prevalence of stomatocyes and multilobe cells is found, while at high shear rates, flattened RBCs are mainly observed. Stomatocytes and polylobes found for moderate $\dot{\gamma}$ values show a tumbling-like dynamics. The dynamics of flattened RBCs at high shear rates resembles a tank-treading motion; however, these cells display persistent and dynamic membrane deformations represented by small creases or larger lobes, which continuously form and disappear (see Movie S7). Tank-treading with a steady ellipsoidal shape was never observed in simulations and the motion of flattened RBCs at high shear rates is likely to be a superposition of both cell stretching and lobes rotation.            

Finally, to highlight the importance of $\lambda$ for the presence of lobular shapes and their role in shear-thinning in comparison to 
steady tank-treading, we perform rheological measurements for suspensions of normal RBCs dispersed in solutions with increasing viscosities modulated by various dextran concentrations of $2\%$, $4\%$ and $9\%$ (wt/wt). Hematocrit is kept at $45\%$ and the suspensions are sheared at 25$^{\circ}$C for the range of $\dot{\gamma}$ between $10$ s$^{-1}$ and $2000$ s$^{-1}$ (without hardening). Figure \ref{fig4}B presents relative viscosities of these suspensions as a function of the shear stress $\tau$ ($=\eta_{s}\dot{\gamma}$). The curves for PBS/BSA and $9\%$ dextran solution display inconsistent trends. In PBS/BSA, which corresponds to high viscosity ratios ($\lambda\approx5-6$), a significant and continuous decrease of $\eta_{rel}$ is observed for increasing $\tau$ up to the rheometer limit of about $6$ Pa. In contrast, $9\%$ dextran 
suspensions ($\lambda\approx0.15$) exhibit a slight shear-thinning for stresses close to $1$ Pa, but $\eta_{rel}$ yields a plateau for 
higher stresses (see fig. \ref{fig4}B). Samples with $2\%$ and $4\%$ dextran solutions show an intermediate behavior, in which the occurrence of an increasingly significant shear-thinning for increasing $\lambda$ is detected. In parallel, we have performed microfluidic observations of single RBCs flowing in $2\%$ and $9\%$ dextran solutions at $Ht=1\%$ for comparable shear stresses. Contrary to the behavior of RBCs in PBS/BSA, cells in $9\%$ dextran solutions do not assume polylobed shapes, but show a rather sharp transition from rigid tumbling discocytes to tank-treading ellipsoids which elongate with increasing stresses. This transition takes place at $\tau\sim 1 Pa$ which is consistent with the change in slope of the corresponding $\eta_{rel}$ curve in fig. \ref{fig4}B. In the inset of fig. \ref{fig4}B an illustration of the effect of $\lambda$ on RBC dynamics is presented. Images show the shapes of cells dispersed respectively in PBS/BSA, $2\%$ and $9\%$ dextran solutions and flowing in a microchannel with the same estimated local shear stress of $\tau=6$ Pa. A decrease in $\lambda$ promotes in-flow orientation, elongation and simultaneous 
reduction of lobes size (see Movie S8). When $\lambda\approx0.15$, polylobed cells are completely replaced by elongated tank-treading ellipsoids (see Inset fig. \ref{fig4}B).
\textcolor{red}{Noteworthy, all images in the inset of fig. \ref{fig4}B and movies in the Supplemental Materials refer to RBCs flowing in microchannels in which the presence of a gradient in $\dot{\gamma}$ can potentially affect RBCs dynamics, while rheological measurements were performed under pure shear flow conditions. However, comparison between off-center shape sequences in 50 $\mu$m capillaries and simple-shear YALES2BIO simulations presented in fig. 2A indicates a rather marginal effect of the gradient of shear rate.}. 

\section{Discussion}

Investigation of blood rheology finds its cornerstone in the seminal work of Chien and collaborators \cite{Chien:1970wz}. These authors have shown how the deformability of RBCs influences blood viscosity, especially at high shear rates and hematocrits. ``Deformability'' means that cells can change their shape in response to mechanical stresses. This term encompasses both out-of-plane and in-plane deformations of membrane elements. In the 1970s, deformability has been mainly attributed to tank-treading observed when RBCs were suspended in a viscous fluid mimicking the viscosity of WB at high shear rates. The resulting stabilization of cells' orientation implied a gradual elongation of the cells for increasing shear rates, which was believed to be the main reason for shear-thinning. This droplet-like behavior of RBCs led to the analogy between blood and emulsion rheology \cite{SchmidSchoenbein:1969tz,Fischer:1978wb}.

Our morphological and rheological studies in combination with microfluidic observations and simulations lead to a different picture. Without aggregation, blood continuous shear-thinning is the result of a rich dynamical behavior of RBCs population, which is based primarily on the inability of RBCs to tank-tread under physiological conditions ($\lambda\approx5$). We have observed a diversity of shapes and dynamical states of RBCs for increasing shear rates and the development of a large fraction of highly deformed and polylobed cells for $\dot{\gamma}>400$ s$^{-1}$ (figs. \ref{fig1} and \ref{fig2}) which has been confirmed by simulations. 

\textcolor{red}{We first discuss single-cell behavior in dilute suspensions.}
For $\dot{\gamma}<1$ s$^{-1}$, the biconcave shape is preserved and the behavior of RBCs is similar to that of a rigid oblate ellipsoid, such that
the membrane and the enclosed fluid rotate as a rigid body. For increasing shear rates up to 10 s$^{-1}$, more and more cells are found to roll
on their edge as previously observed in experiments using rheoscopes \cite{Bitbol:1986tp} and more recently in flow chambers \cite{Yao:2001ux,Dupire:2012cx}.
Shear-thinning in this range of shear rates is therefore mainly controlled by discocyte orientation. Rolling has its origin in shear elasticity of
the membrane. Under physiological conditions, membrane tank-treading is forbidden due to the relatively high internal viscosity as predicted
by viscous ellipsoidal models of the RBC \cite{Skalak:2005wr}. However, during each tumbling period the membrane elements oscillate around a given position and
this local oscillatory strain seems to destabilize RBCs from tumbling toward rolling \cite{Dupire:2015vj} instead of tank-treading.
Though not fully settled, this phenomenon is well captured by our simulations and is described as a stable motion in several recent numerical
simulations of capsules \cite{Dupont:2013go} and RBCs \cite{Cordasco:2013go}. 

In the range of shear rates between 45 and 400 s$^{-1}$, a large population of the rolling discocytes looses one dimple and becomes rolling
stomatocytes displaying at first more frequently a circular rim up to 45 s$^{-1}$ and then a breathing elliptical rim for shear rates reaching 400 s$^{-1}$.
In this second morphological transition, rolling stomatocytes are showing a smaller cross-section to shear flow than rolling discocytes,
reducing their disturbances of the flow lines. This leads to a further shear-thinning.  Both our numerical approaches capture this flow-induced morphological transition
for single RBCs. The loss of the dimple happens abruptly in simulations highlighting a buckling instability, which is also confirmed experimentally
by the long time scales (minutes) necessary for the population of stomatocytes to relax to a discocyte shape. 

Beyond 400 s$^{-1}$, the dilute population of RBCs is subjected to a new shift in morphology and large lobes develop on the surface of the cells (fig. \ref{fig1}A).
These out-of-plane deformations are again a signature for a lack of fluidization of the membrane even for the largest stress of 6 Pa we were able to reach
both experimentally (with the rheometer and in microflows) and numerically. Noteworthy, Sutera and Mehrjardi \cite{Sutera:1975}  studied in the 1970s the morphology
of cells hardened in PBS solutions at very high shear stresses ($\tau\geq10$ Pa). Their images at $\tau=10$ Pa show the presence of pronounced concavities
on RBCs similar to those we observe for polylobed cells. Thus, trilobes or even hexalobes are the result of the large viscosity contrast. Our experiments
with varying viscous environment confirm how lobes are directly associated with a large enough viscosity contrast $\lambda$ since their presence vanishes in very viscous dextran
solutions as shown in fig. \ref{fig4}. However, the exact nature of these transitions and the development of folded shapes remain to be understood.
Both experiments and simulations suggest that large enough viscosity ratio is at least sufficient for polylobes to appear. However, modifications
 of a RBC membrane at rest can also induce similar transformations observed here for increasing shear rates. Theoretically, a large equilibrium-shape variety has been 
obtained for RBCs including stomatocytes, with circular and deformed rims, and trilobes (named knizocytes in the literature), 
by changing the resting elastic state of the cytoskeleton and the area difference between lipid leaflets in their model \cite{Lim:2002}. 
Despite this striking similarity, the dynamic shapes in simulations under flow have been
observed without any explicit modification of the membrane. These results open the road for further experimental investigations to explore
the effect of flow on the shape modification of a RBC membrane.  

Our dilute single-cell experiments and simulations show how out-of-plane deformations are inevitable for increasing shear rates. However, \textcolor{red}{for more concentrated suspensions and as hematocrit
is increased}, a large population of creased and flattened discocytes appears both in experiments (see Movie S9) and simulations (see for instance Movie S7 in supplementary information
and fig. \ref{fig3}B). While it is perfectly conceivable that large lobes cannot easily develop on membranes in crowded suspensions, the complex flow
between cells seems to counter-balance the inherent tendency of RBCs to elongate more in the vorticity direction. It is this fine balance between vorticity-
and flow-direction elongations that aligns cells on average parallel to the flow. Nevertheless, any variations in relative positions between the cells or
any collisions immediately give birth to local bulges on the membranes and produce large fluctuations of shapes as seen in RBCs extension distributions of the simulation presented in fig. \ref{fig3}B. 

In conclusion, our study reveals that blood shear-thinning is related to a rich behavior of RBCs in shear flow convoluted with a large distribution
of cell shapes for any given flow condition.
\textcolor[rgb]{1,0,0}{The lack of membrane fluidity for high viscosity contrast between inner and outer fluids is the key feature which controls RBCs behavior.
As a consequence, the droplet-based analogy for blood rheology at high shear rates appears to be erroneous for the explanation of shear-thinning. Moreover, several fundamental physiological phenomena have been analyzed under the assumption of membrane tank-treading, such as vaso-regulatory ATP release by RBCs in strong shear flows \cite{Forsyth:2011jx} or the formation of a few-micron thick cell-free layer adjacent to the vessels walls which is responsible for the apparent viscosity drop with decreasing vessel diameter (the so-called F\aa hr\ae us-Lindqvist effect \cite{Fahraeus:1931}). Our study questions the relevance of a droplet-like analogy for RBC dynamics to explain these phenomena and asks to re-examine them both experimentally and theoretically for physiologically relevant viscosity and stress conditions. Finally, our study suggests that in pathological change in plasma composition, RBCs cytosol viscosity or membrane mechanical properties, will impact  the onset of shape transitions and should play a central role in pathological blood rheology and flow behavior.} 



\section{Materials and Methods}
  \subsection{Sample preparation}
  Fresh venous-blood \textcolor[rgb]{1,0,0}{samples were obtained} from healthy consenting donors, \textcolor[rgb]{1,0,0}{thanks to the agreement with a local blood
    bank to use blood (EFS) and its components for non-therapeutic purposes}. \textcolor[rgb]{1,0,0}{The samples, whose volume ranged between 420 and 500 ml, were stored in
    PVC-DEHP transfusion bags produced by Macopharma (MSE Systems). Such bags are provided with 66.5 ml of CPD anticoagulant, whose composition is: Sodium citrate dihydrate
    26.3 g/l, Citric acid monohydrate 3.27 g /l, Monobasic Sodium Phosphate dihydrate 2.51 g/l, Dextrose monohydrate 25.5 g/l}. The suspensions were then centrifuged at
  1000 x g for 20 min and the buffy coat was removed. RBC \textcolor[rgb]{1,0,0}{dispersions} were diluted in PBS and the procedure of centrifugation/removal was repeated
  twice to obtain washed RBCs. Afterwards, cells were re-suspended in their native plasma \textcolor[rgb]{1,0,0}{treated with anticoagulant} or in solutions of PBS
  supplemented with BSA (1 mg of BSA per 1 ml PBS) and the hematocrit was adjusted by using graduated and narrow test-tubes. Samples were kept at 25$^{\circ}$C throughout all the preparation. 
	
Dilute suspensions ($Ht=1$\%) of fluorescent HRBCs for confocal observation were prepared by using a PKH26 Red Fluorescent Cell Linker (Sigma-Aldrich). HRBCs were labeled following standard protocol of the provider. 
	
\textcolor[rgb]{1,0,0}{2,000 kDa Dextran (Leuconostoc spp., Sigma-Aldrich) solutions were prepared in PBS with a concentration of 2\% (wt/wt), 4\% (wt/wt), and 9\% (wt/wt).
  Dissolution was obtained at room temperature by overnight gentle stirring. All dextran/PBS solutions showed a nearly Newtonian behavior with a constant viscosity equal to 3.4,
  10.2 and 38.8 mPa*s respectively for dextran concentrations of 2\% (wt/wt), 4\% (wt/wt), and 9\% (wt/wt). We used a very high molecular weight dextran in order not to change
  the osmolarity of the buffer suspension when it has been added to make the suspending medium more viscous. 
  All dispersions containing RBCs and all suspending solutions used in our experiments have shown a pH ranging between 7.2 and 7.5 and osmolarities of our buffers were measured
  to be 285 mOsm on average. The samples were stored in a refrigerator at 4$^{\circ}$C and they were used within the time frame between 3 hours and 5 days after the blood draw. The healthy state of
  the cells have been checked by looking at the amount of echinocytes before each experiments}. (Blood samples and realized experiments are summarized in table 1 of the supplementary text).

  \subsection{Rheology and microfluidics}
  Rheology experiments were performed in a cone-plate stress-imposed rheometer (AR 2000 - TA Instrument). \textcolor{red}{The temperature was kept constant at $37^\circ$C by a Peltier temperature controller. Data were analyzed by the Rheology Advantage Data Analysis software. In the first type of experiments, diluted suspensions of RBCs (Ht=1\%) in a PBS/BSA solution were hardened in flow by adding a 70\% wt/wt solution of glutaraldehyde prepared in PBS solution (Sigma-Aldrich) and glutaraldehyde concentration was adjusted to 9\%. 300 $\mu$l of the resulting sample was added by capillarity to 700 $\mu$l of the RBC dispersion during shearing to reach a final glutaraldehyde concentration of almost 3\%. Such an amount of glutaraldehyde produces a very rapid, strong and permanent solidification. Indeed, recent AFM indentation measurements \cite{Dulinska:2006} performed on glutaraldehyde-treated RBCs have yielded a Young's modulus $E$ in the range 10-100 kPa. Therefore, for the maximum shear stress $\sigma$ of 6 Pa reached in our experiments, the typical strain of hardened RBCs, $\sigma/E$, is expected to be within 0.05-0.005\%. This means that HRBCs can be considered as rigid objects in all our experiments}. The same protocol was used to harden samples at higher hematocrits. Both shearing and hardening were carried out within 2 min. \textcolor[rgb]{1,0,0}{Samples were then collected, washed, and suspended in PBS solution for microscopic examination.}

  Images of stiffened cells were acquired with an inverted microscope (Olympus IX71 60x objective) equipped with a digital camera (Sony XCD-X710). Acquired
  data were analyzed in Matlab (Mathworks,
version 2011a): the minimum number of cells counted and classified for each $\dot{\gamma}$ in fig. \ref{fig1}B and for each $Ht$ in
  fig. \ref{fig3}A was equal to 550 and 450, respectively. Microflow observations were realized at 25$^{\circ}$C in glass round capillaries with a diameter
  of 50 $\mu$m. Flow rates were adjusted by a OB1 Pressure Controller (ElveFlow) operating up to 2 bar. Images were acquired at a high magnification
  (60x and 100x objectives) on a Nikon inverted microscope equipped with a high-speed camera Phantom Miro M320S (Vision Research) and subsequently analyzed
  by a custom image analysis software (ImageJ).

\subsection{Numerical simulations}
Two different in-house softwares were used for numerical simulations. The first one, YALES2BIO \cite{Mendez:2014a,Gibaud:2015},
 is a finite-volume parallel solver for the incompressible Navier-Stokes equations on unstructured meshes. Fluid-structure coupling was implemented using an immersed boundary
method adapted to unstructured grids \cite{Mendez:2014a}. RBCs were modeled as viscous drops enclosed by membranes resisting shear, bending, and area dilation. 
In the SDPD simulations, the smoothed dissipative particle dynamics method \cite{Mueller_SDPD_2015}, a mesoscale hydrodynamic particle-based approach, represents fluid flow, while a RBC membrane was modeled as a triangulated network
of springs \cite{Fedosov_RBC_2010}, whose vertices are coupled to the fluid via frictional forces. The network assumes fixed connectivity and includes the spring's elastic energy,
bending energy, and area and volume conservation constraints \cite{Fedosov_RBC_2010}. Further details about the numerical methods and setups can be found in supplementary information.




\section*{Acknowledgments}
  M.A., S.M., F.N. and L.L. acknowledge the support of the OSEO-BPI project Dat\symbol{64}Diag. M.A. and L.L. are grateful for the support from the Labex Numev (convention No. ANR-10-LABX-20).
  S.M. and  F.N. acknowledge the support of the ANR (ANR-11-JS09-0011) and thank Dr. Moureau and Dr. Lartigue (CORIA, UMR 6614) and the SUCCESS scientific group for providing
  the YALES2 code, which served as a basis for the development of YALES2BIO. Simulations with YALES2BIO were performed using HPC resources from GENCI-CINES (Grant  2015-c2015037194).
  D.A.F. acknowledges funding by the Alexander von Humboldt Foundation. SDPD simulations were performed using a CPU time allocation at the J\"ulich Supercomputing Center.






\bibliographystyle{pnas}
	
\clearpage

\subsection*{Legends}

\begin{figure}[h]
\caption{Investigation of RBC shapes at different shear rates in a cone-and-plate rheometer. A) Observation of hardened cells by optical (black and white) and
  confocal (red) microscopy: with increasing $\dot{\gamma}$ the formation of highly deformed stomatocytes and then polylobed cells \textcolor{red}{(trilobe and hexalobe)} are detected
  (scale bars are 5 ${\mu}$m). B) Shape distribution of RBCs populations in samples hardened at different shear rates: the three regions in color highlight different regime of decrease of discocytes population. \textcolor{red}{The error bars represent triplicate measurements realized in the two rapidly varying regions. They illustrate the typical variance of the measurements.}}
\label{fig1}
\end{figure}
\newpage

\clearpage

\begin{figure}[h]
\caption{ \textcolor[rgb]{1,0,0}{Microfluidic observations of RBC dynamics in shear flow. A) Time lapse sequence of deformation of RBCs at different $\dot{\gamma}$: Time lapses of 20 ms for the tumbling discocyte, 6 ms for the rolling discocyte, 3.25 ms for the 3 stomatocytes, 0.6 ms for the trilobe in top view, 1.75 ms for the trilobe in side view and 0.6 ms for the hexalobe. Right panel corresponds to analogous time sequences of RBCs obtained with YALES2BIO simulations: Time lapses are given in $1/\dot{\gamma}$ units: $8$ for the tumbling discocyte, $7$ for the rolling discocyte, $6$ for the stomatocyte and the tumbling deformed stomatocyte, and $7$ for the four last cases. B) Stop flow sequences of (left) a trilobed and (right) an hexlobed cell. For the trilobe, the total relaxation time is 1 s and the intermediate images are separated by respectively 0.23 and 0.56 s. For the hexalobe, relaxation occurs in 1.18 s and successive images have a time interval of respectively 0.32 and 0.71 s. (scale bars are 5 ${\mu}$m).}. 
}
\label{fig2}
\end{figure}

\clearpage

\clearpage

\begin{figure}[h]
\caption{\textcolor[rgb]{1,0,0}{Characterization of RBCs morphology as a function of increasing $Ht$. A) Shape distributions of HRBCs at 900 s$^{-1}$ in suspensions with different $Ht$. For the sake of clarity, the number densities of discocytes and stomatocytes, which are negligible, have been omitted. Inset: black frames on the right, a top view and a cross section of a creased discocyte acquired by confocal microscopy as well as an image of a polylobed irregular shape (multilobe) obtained by optical microscopy (scale bar is 5 ${\mu}$m). Physiological hematocrits in 
the microcirculation lie in the range $15-30$\% indicated by the light yellow region. B) Distributions of RBCs extension for different $Ht$ values measured in SDPD simulations. The extension is defined as the maximum instantaneous cell length in the flow direction, as depicted in one of the insets. The distributions include the data from multiple RBCs in a suspension and are averaged over many time instances. The insets illustrate typical shapes for different values of the cell extension. The two extension regions marked by light yellow and light blue colors schematically depict cell-extension ranges, where polylobed shapes and creased discocytes are observed.}
}
\label{fig3}
\end{figure}

\clearpage

\clearpage

\begin{figure}[h]
\caption{Rheology of dense suspensions of RBCs ($Ht=45\%$). A) Relative viscosity of the suspensions of deformable RBCs in plasma (red) and in PBS/BSA 
(yellow) as a function of $\dot{\gamma}$ in comparison to the suspensions of washed cells hardened at rest (green) and at $\dot{\gamma}$=1500 s$^{-1}$ 
(blue). Suspensions with a deformable corpuscular phase show a typical shear-thinning for increasing $\dot{\gamma}$, whereas hardened samples have 
  a nearly Newtonian behavior. \textcolor{red}{SDPD simulation data (black stars) are also shown for deformable cells and agree well with the experimental results for RBCs suspended in plasma or in PBS/BSA.}
  B) Rheology of washed blood in comparison to the suspensions of RBCs in solutions with different dextran concentrations 
(T=25 $^{\circ}$C). The effect of viscosity ratio $\lambda$ is highlighted in the inset by the illustration of single cells flowing in both PBS/BSA and dextran 
solutions in microfluidics at a comparable shear stress $\tau$. The scale bars are equal to 5 $\mu$m.}
\label{fig4}
\end{figure}

\end{document}